# Dynamic Tilting of Ferroelectric Domain Walls via Optically Induced Electronic Screening

Youngjun Ahn,[1] Arnoud S. Everhardt,[2] Hyeon Jun Lee,[1] Joonkyu Park,[1] Anastasios Pateras,[1] Silvia Damerio,[2] Tao Zhou,[3] Anthony D. DiChiara,[4] Haidan Wen,[4] Beatriz Noheda,[2,5] and Paul G. Evans[1,‡]

[1] *Department of Materials Science and Engineering, University of Wisconsin-Madison, Madison, Wisconsin 53706, USA*

[2] *Zernike Institute for Advanced Materials, University of Groningen, 9747AG- Groningen, Netherlands*

[3] *ID01/ESRF, 71 Avenue des Martyrs, 38000 Grenoble Cedex, France*

[4] *Advanced Photon Source, Argonne National Laboratory, Argonne, Illinois 60439, USA*

[5] *CogniGron Center, University of Groningen, 9747AG- Groningen, Netherlands*

‡pgevans@wisc.edu

## Abstract

Optical excitation perturbs the balance of phenomena selecting the tilt orientation of domain walls within ferroelectric thin films. The high carrier density induced in a low-strain $BaTiO_3$ thin film by an above-bandgap ultrafast optical pulse changes the tilt angle that 90° *a*/*c* domain walls form with respect to the substrate-film interface. The dynamics of the changes are apparent in time-resolved synchrotron x-ray scattering studies of the domain diffuse scattering. Tilting occurs at 298 K, a temperature at which the *a*/*b* and *a*/*c* domain phases coexist but is absent at 343 K in the better ordered single-phase *a*/*c* regime. Phase coexistence at 298 K leads to increased domain-wall charge density, and thus a larger screening effect than in the single-phase regime. The screening mechanism points to new directions for the manipulation of nanoscale ferroelectricity.

Nanoscale-thickness regions near ferroelectric domain walls exhibit properties that are distinct from bulk or thin-film ferroelectric materials due to the rapid spatial variation of the polarization and inhomogeneous distortion of the crystal lattice [1,2]. The dramatically different electronic properties of domain-wall regions arise largely from two effects. First, a step in the electrostatic potential and a related high density of bound charges can arise at domain walls [3,4]. In addition, inhomogeneous strain or oxygen octahedral rotation near domain walls can modify the band structure via the deformation potential and flexoelectric coupling [5-7]. These effects can lead, for example, to room-temperature electrical conductivity in otherwise insulating ferroelectrics and to the formation of a two-dimensional electron gas [8-10]. The electrostatic potential step can separate photoexcited electron-hole pairs and contribute to the generation of photovoltages larger than the electronic bandgap [3,11-13]. The novel phenomena associated with domain walls make it important to probe how the physical properties of nanoscale volumes near domain walls evolve under external stimuli. In this Letter, we show that elastic heterogeneity introduced by a coexistence of two different types of ferroelastic domain patterns leads to unusual nanosecond-timescale responses of domain wall configuration to femtosecond-duration optical pulses.

The orientation of domain walls with respect to crystallographic directions, surfaces, or interfaces depends on multiple contributions to the free energy, including stress and the density of bound charges [2,14]. In addition to these unit-cell-scale effects, the bound charge density can also arise from mesoscopic geometric deviations from the ideal domain configuration such as domain-wall roughness [7,14]. The dependence of the bound charge density on this disorder in the domain configuration leads to an intriguing link between electronic or optoelectronic effects and the domain-pattern phase diagram.

We report an optically induced modulation of the electrostatic energy contribution to the domain wall energy. The experimental signature of this modulation is a change in the angle formed between the plane of domain walls in a $BaTiO_3$ (BTO) thin film and the surface, termed the tilt angle $\alpha$. The change in $\alpha$ occurs on the single-nanosecond timescale following excitation by an ultrafast optical pulse and relaxes over several nanoseconds. Changes in the electrostatic contribution to the free energy can occur through screening of the bound charge by mobile charges, including those created by optical absorption [2]. The origin of the bound charges is apparent in the temperature dependence of the experimental observation. The tilting occurs at a temperature at which there is a coexistence of multiple domain configurations and a resulting deviation from the ideal zero-bound-charge domain arrangement.

Experiments probing the optically induced domain wall tilting used an epitaxial 78 nm-thick BTO film on a 6 nm-thick $SrRuO_3$ layer on an $NdScO_3$ (NSO) substrate, as in Fig. 1(a) [15]. The scattered x-ray intensity was analyzed using reciprocal-space coordinates $Q_x$, $Q_y$, and $Q_z$ along the [100], [010], and [001] pseudo-cubic (pc) directions, which correspond to [-110], [001], and [110] orthorhombic (o) directions in NSO, respectively. Directions are given here with subscripts indicating the basis.

In the temperature range probed in these experiments, the BTO layer can be generally described by means of two domain configurations: (i) a high-temperature *a*/*c* pattern with alternating orthogonal domains of in-plane and out-of-plane polarization, and (ii) an *a*/*b* pattern appearing below 340 K with stripes of orthogonal in-plane polarization. The *a*/*c* domain pattern has period $\Lambda$ along $[100]_{pc}$ and equilibrium domain-wall tilt angle $\alpha_0$, as illustrated in Fig. 1(a) [15]. In addition to the *a*- and *c*-components of the polarization in Fig. 1(a), there are indications that the *a*/*c* pattern at 298 K also includes a small in-plane component *a** along $[010]_{pc}$ [16,17], such that the 298 K domain phase would be the predicted *aa**/*ca** configuration [16]. The *a** component is predicted to exhibit a 180° rotation at the *a*/*c* domain wall. Temperature-dependent x-ray diffraction measurements described in the Supplementary Materials and piezoelectric force microscopy both indicate that *a*/*c* and *a*/*b* phases coexist over a temperature range of tens of degrees below 340 K [18,19]. At 298 K, the majority of the volume of the BTO film is in the *a*/*b* phase. The transition from ferroelectric to paraelectric phases occurs at 403 K which is outside the temperature range of the experiments reported here [15].

Time-resolved synchrotron x-ray diffraction experiments were performed at station 7-ID-C of the Advanced Photon Source. The x-ray photon energy and pulse duration were 9 keV and 100 ps, respectively. The scattered intensity was measured in the three-dimensional volume of reciprocal space near the 002 BTO reflection. Femtosecond-duration laser pulses with 400 nm-wavelength (optical photon energy $\hbar\omega$=3.1 eV) and absorbed fluence $F_{abs}$ were synchronized with the x-ray pulses with variable delay $t$. The step size of the delay time used for the experiments ranges from 100 ps to 2 ns. Further details are in the Supplementary Materials [20].

The x-ray intensity in a $Q_x$-$Q_z$ section of reciprocal space at $Q_y = 0$ near the 002 BTO reflection is shown in Fig. 1(b). A streak of intensity arising from the *a*/*c* domain pattern extends from high to low $Q_z$ with increasing $Q_x$, forming an angle $\alpha_0$=42°. Second-order diffraction intensity maxima at $Q_x = \pm 4\pi/\Lambda$ are apparent in the inset above Fig. 1(b). The intensity maxima have a separation of $2\pi/\Lambda$=0.008 Å$^{-1}$ along $Q_x$ at 298 K with $\Lambda$=78 nm. Domain scattering maxima are indexed with orders -2, -1, +1, and +2 such that maxima with negative $Q_x$ have a negative order.

Optical excitation leads to an out-of-plane expansion of the BTO lattice parameter and to a change in $\alpha$. The effect used to measure time-dependence of the tilt angle $\alpha(t)$ is illustrated in Fig. 1(c) using the $Q_z$ profiles of the -1 and +1 domain scattering at $t$=1 ns for $F_{abs}$ = 2.4 $\mu$J/cm$^2$. The intensity maxima of the -1 and +1 orders exhibit different optically induced fractional shifts of $Q_z$:

-0.006% and -0.015%, respectively. The different shifts of the two orders indicate that there is a change in the angle of the domain scattering streak. At 1 ns the change is $\Delta\alpha(t=1 \text{ ns}) = -0.5°$ where $\Delta\alpha(t) = \alpha(t)-\alpha_0$. The negative value of $\Delta\alpha$ indicates that the angle that the domain wall forms with the substrate surface is slightly reduced. The area of the domain wall thus increases following optical excitation, as described in more detail below. The peak positions of the ±1 orders of domain scattering along $Q_x$ remain unchanged within the experimental uncertainty.

The time dependence of the distribution of scattered intensity along $Q_z$ is shown for the ±1 and ±2 orders of the domain scattering in Figs. 2(a) to (d) for $F_{abs} = 2.4\ \mu\text{J/cm}^2$. The difference in the shifts of wavevectors of the intensity maxima of the ±2 orders are consistent with changes in $\alpha$. The intensity variation observed in Fig. 2 can, in principle, arise from the photo-induced changes in magnitude of the ferroelectric polarization, either at the domain walls or in the remaining volume of the film [21-23]. The polarization depends on the off-centering of Ti ions within the oxygen octahedra [24], on which the diffracted intensity in turn depends through the structure factor. The oscillation of the intensity appears to be an experimental artifact associated with a combination of the different discrete time steps used at different stages in the measurements in Fig. 2 and experimental uncertainty. Detailed discussion of the intensity changes, however, is outside the scope of this Letter.

The dynamics of the fractional changes in the maximum $Q_z$ of the ±1 orders of the domain scattering and the resulting $\Delta\alpha(t)$ are shown in Fig. 3(a) for $F_{abs} = 2.4\ \mu\text{J/cm}^2$. $\Delta\alpha$ changes by -0.5° within 1 ns after excitation, followed by relaxation over tens of ns. The relaxation time is similar to the time constant for recombination of photoexcited carriers [25]. The sign of $\Delta\alpha$ is different from the shift expected due to heating because (i) heating from 298 K induces positive change in $\alpha$ and (ii) the change in tilt angle due to heating has a non-monotonic temperature dependence, as described in the Supplementary Materials [20].

The tilting is not observed in scattering measurements conducted in the single-phase *a*/*c* regime at 343 K, providing an indication that the tilt is linked to features of the domain configuration. Figure 3(b) shows the time dependence of $Q_z$ at intensity maxima of the +1 and −1 orders of domain scattering and the value of $\Delta\alpha(t)$ at 343 K for $F_{abs}$ = 6.5 $\mu$J/cm$^2$. The intensity at both orders shifts by $\Delta Q_z/Q_z$ by -0.015% at $t$=1 ns, indicating that the lattice expansion is optically induced at 343 K, as at lower temperatures. The magnitudes of the shifts at both orders are equal, however, which reveals that $\Delta\alpha$=0 at 343 K, even with higher optical fluence than at 298 K. The optical fluence dependences of $\Delta\alpha(t$=1 ns) at 298 K and 343 K are shown in Fig. 3(c). The magnitude of $\Delta\alpha(t$=1 ns) increases as a function of $F_{abs}$ at 298 K, reaching -0.85° at 5.0 $\mu$J/cm$^2$. At 343 K, $\Delta\alpha(t$=1 ns) is zero within experimental uncertainty for optical fluences up to 15 $\mu$J/cm$^2$.

A model based on screening by photoinduced charge carriers accurately accounts for the observed tilting phenomenon and its dependence on experimental parameters. In the absence of bound charge, the domain pattern adopts a tilt orientation that minimizes the elastic energy. The head-to-tail orientation of the polarization in the *a*/*c* or *a*/*b* patterns nominally ensures that there is no net bound charge at the domain walls. Charged domain walls may, however, exist in BTO thin films due to deviations in the domain wall angle or the mesoscopic arrangement of domains walls, inhomogeneous stress on domain walls, and roughness of the domain walls [2,7,26]. Domain-wall roughening can arise from impurities, dislocations, and local strain gradients [27,28]. The domain-wall roughness can lead to the accumulation of bound charge at the small fraction of sites at which the polarization does not fulfil the local polarization continuity condition [14]. The magnitude of the bound charge density is much lower than would arise from the discontinuity of the total remnant polarization $P_0$ and varies significantly depending on the domain configuration [7].

Reported charge densities at roughened domain boundaries are on the order of 10% or less of the strongly charged case, a magnitude that can be screened by the photoinduced charge densities here [2,26]. The formation of bound charges may also arise from a discontinuity of the predicted *a** polarization component of the *a*/*c* phase. In this case, bound charge of a fraction of the predicted few-$\mu$C cm$^{-2}$ magnitude of the *a** component would arise in regions in which the in-plane direction of the domain walls is not aligned with the nominal $[010]_{pc}$ direction or at the boundaries between the *a*/*c* and *a*/*b* phases [15,16].

The substantial electrostatic energy per unit area of domain walls at which there is a non-zero density of bound charges causes the equilibrium configuration of domain walls to tilt towards the film surface with respect to the elastically preferred orientation in order to reduce the area of the domain wall and thus the charge density. Optically excited carriers screen the bound charges, leading to a reduction in electrostatic energy associated with charged domain walls and to a tilting of domain walls towards the substrate, as a result of which the domain wall area increases. The charge density induced by optically excited carriers for the experimental flux employed in Fig. 2 is $F_{abs}/\hbar\omega = 5 \times 10^{12}$ cm$^{-2}$, assuming that each absorbed photon produces one excited carrier. The optically induced charged density is orders of magnitude smaller than the bound charge density at strongly charged domain boundaries, which is on the order of $P_0$, the equivalent of $10^{14}$ cm$^{-2}$. The photoinduced charge density would thus not be sufficient to screen a strongly charged domain wall. The BTO thin films considered here, however, have weakly charged walls with far smaller charge densities consistent with geometries that result in a nearly continuous polarization. In each possibility, namely roughness/disorder or the discontinuity of an *a** component, the magnitude of the bound charge is similar to the photoinduced charge density available for screening.

The disorder and coexistence of the domain patterns, and the link to a larger bound charge density at 298 K, is apparent in the intensity distribution in $Q_x$-$Q_y$ sections of reciprocal space, revealing the in-plane ordering of the domain patterns. The scattered intensity in $Q_x$-$Q_y$ sections at 298 K and 343 K is shown in Figs. 4(a) and (b), respectively. Scattering from the *a*/*c* pattern appears along the line at $Q_y$=0 at both temperatures. Figure 4(a), acquired at 298 K, also has two additional pairs of intensity maxima distributed diagonally along $<110>_{pc}$ directions arising from the *a*/*b* domain pattern. There are several differences in the scattering pattern in the configuration consisting of only the *a*/*c* phase at 343 K. The *a*/*c* pattern diffuse scattering exhibits a higher overall intensity and stronger ±2 order maxima at 343 K due to the absence of the *a*/*b* pattern at this temperature.

The coexistence of the domain phases at 298 K leads to a structural distortion that is apparent in the reciprocal-space widths of the ±1 orders of *a*/*c* domain scattering. The widths of the domain scattering intensity along $Q_x$ and $Q_y$, shown in Figs. 4(c) and (d), are far larger than the natural width of the structural reflections because the domain period is coherent over a finite in-plane distance [29]. The in-plane coherence lengths along $[100]_{pc}$ and $[010]_{pc}$ calculated from the full-width-at-half-maximum (FWHM) of the domain scattering are 140 nm and 700 nm, respectively, at 298 K. The coherence lengths are 20% larger along both directions at 343 K than at 298 K. Coherence lengths measured using x-ray scattering include the effects of subtle variations in strain and domain period and are thus smaller than the size of domain phase regions observed using piezoelectric force microscopy [15]. The increased coherence length of the domain pattern at elevated temperature is somewhat counterintuitive because the roughness of the domain walls and domain disorder generally increase as a function of temperature [27,30]. In this case, however, the comparatively large widths at 298 K are consistent with the coexistence and disorder of the domain phase.

We can also consider other mechanisms, besides the optically induced screening, that could potentially contribute to a change in the tilt angle of the domain walls. Optical absorption can lead both to photoinduced elastic strain in several ferroelectric materials [25,31] and, separately, to a reduction of the width of the *a* domain within the repeating unit of the striped *a*/*c* domain pattern [21]. To examine the change in $\alpha$ as a function of out-of-plane strain and *a*-domain width, we used thermodynamic models in which a continuous distribution of infinitesimal edge dislocations is employed to evaluate the energy associated with the elastic compatibility at domain walls. A detailed description of the models is provided in the Supplementary Materials [20]. Briefly, the domain-wall free energy density was computed as a function of $\alpha$ and the width of the *a* domain component, including elastic contributions. The thermodynamic calculations show that a lattice expansion of 0.01%, as observed for $F_{abs}$ = 2.4 $\mu$J/cm$^2$, leads to $\Delta\alpha$ = 0.001°, towards the substrate normal, a change with a far smaller magnitude and opposite sign to the observed effect. Similarly, a reduction of the width of the *a*-domain could lead to a change in the tilt but by an order of magnitude lower than the measured $\Delta\alpha$. Neither of the alternative mechanisms besides screening is quantitatively consistent with the experimental observations.

The time to reach the maximum tilt decreased from 1.2 ns to 0.4 ns as the absorbed fluence increased in the range from 2.4 to 5 $\mu$J/cm$^2$, as shown in the Supplementary Materials [20]. The 1 ns timescale is consistent with a mechanism in which there is a thermodynamic shift in the preferred value of $\alpha$, followed by a kinetic response involving the motion of the domain boundaries to satisfy the new most favorable configuration. The time required for the initial change in $\Delta\alpha$ is compatible with the lateral motion of domain walls to facilitate the transformation. The change in angle requires domain boundaries to move on the order of nm, which is consistent with observed domain wall velocities [32,33].

The time-resolved synchrotron x-ray diffraction experiments reported here reveal an electronic-screening-driven domain wall tilting effect with a nanosecond characteristic timescale. The domain wall tilting uniquely occurs in the domain configuration in which there is elastic heterogeneity near domain walls due to a coexistence of different domain patterns. The mechanism relating the domain tilting to domain wall charging allows the tilting and other domain distortion effects to be used to probe the existence of domain wall charge. The structural heterogeneity in complex domain patterns is, further, a route towards the discovery of unusual responses to external perturbations.

This work was supported by the U.S. National Science Foundation through grant number DMR-1609545. A. S. E., S. D., and B. N. acknowledge financial support from the alumni organization of the University of Groningen, De Aduarderking (Ubbo Emmius Fonds). H. W. acknowledges the support of U.S. Department of Energy, Office of Science, Basic Energy Sciences, Materials Sciences and Engineering Division, for instrumentation development of time-resolved x-ray microdiffraction. This research used resources of the Advanced Photon Source, a U.S. Department of Energy Office of Science User Facility operated for the DOE Office of Science by Argonne National Laboratory under Contract No. DE-AC02-06CH11357. The authors gratefully acknowledge use of facilities and instrumentation supported by NSF through the University of Wisconsin Materials Research Science and Engineering Center, grant number DMR-1720415.

FIG 1. (a) BTO thin film with arrangement and atomic structure of *a*/*c* domain pattern. (b) Scattered x-ray intensity in the $Q_x$-$Q_z$ section of reciprocal space at $Q_y = 0$. The BTO 002 reflection is at $Q_z$=3.13 Å$^{-1}$. Intensity oscillations corresponding to the BTO thickness are along $Q_x = 0$. The intensity distribution along $Q_x$ (inset) is obtained by integrating the intensity with respect to $Q_z$. The ±1 and ±2 orders of domain scattering appear at $Q_x = \pm 0.008$ Å$^{-1}$ and ±0.016 Å$^{-1}$. (c) Intensity profiles of -1 and +1 orders of domain scattering before optical excitation ($t$<0) and at $t$=1 ns for $F_{abs}$ = 2.4 $\mu$J/cm$^2$. The maximum-intensity values of $Q_z$ are indicated with arrows at each time.

FIG 2. Time dependence of the intensities of the (a) -1, (b) +1, (c) -2, and (d) +2 orders of domain scattering as a function of $\Delta Q_z/Q_z$ following optical excitation at $F_{abs}$ = 2.4 $\mu$J/cm$^2$. $\Delta Q_z/Q_z = 0$ corresponds to wavevectors of the intensity maxima before optical excitation. Intensities are normalized to values before optical excitation for each order of domain scattering.

FIG 3. Time dependence of fractional change wavevector $Q_z$ of -1 and +1 orders of domain scattering at (a) 298 K for $F_{abs}$ = 2.4 $\mu$J/cm$^2$ and (b) 343 K for $F_{abs}$ = 6.5 $\mu$J/cm$^2$. (c) $\Delta\alpha$($t$=1 ns) at T = 298 K and 343 K as a function of $F_{abs}$.

FIG 4. Diffracted x-ray intensity distributions in the $Q_x$-$Q_z$ section at (a) 298 K and (b) 343 K. Intensity profiles of -1 and +1 orders of domain scattering along (c) $Q_x$ and (d) along $Q_y$ at 298 and 343 K. The $Q_x$-$Q_y$ sections are obtained by integrating diffracted x-intensity distributions from $Q_z$ = 3.105 to 3.16 Å$^{-1}$.

Ahn *et al*., Figure 1

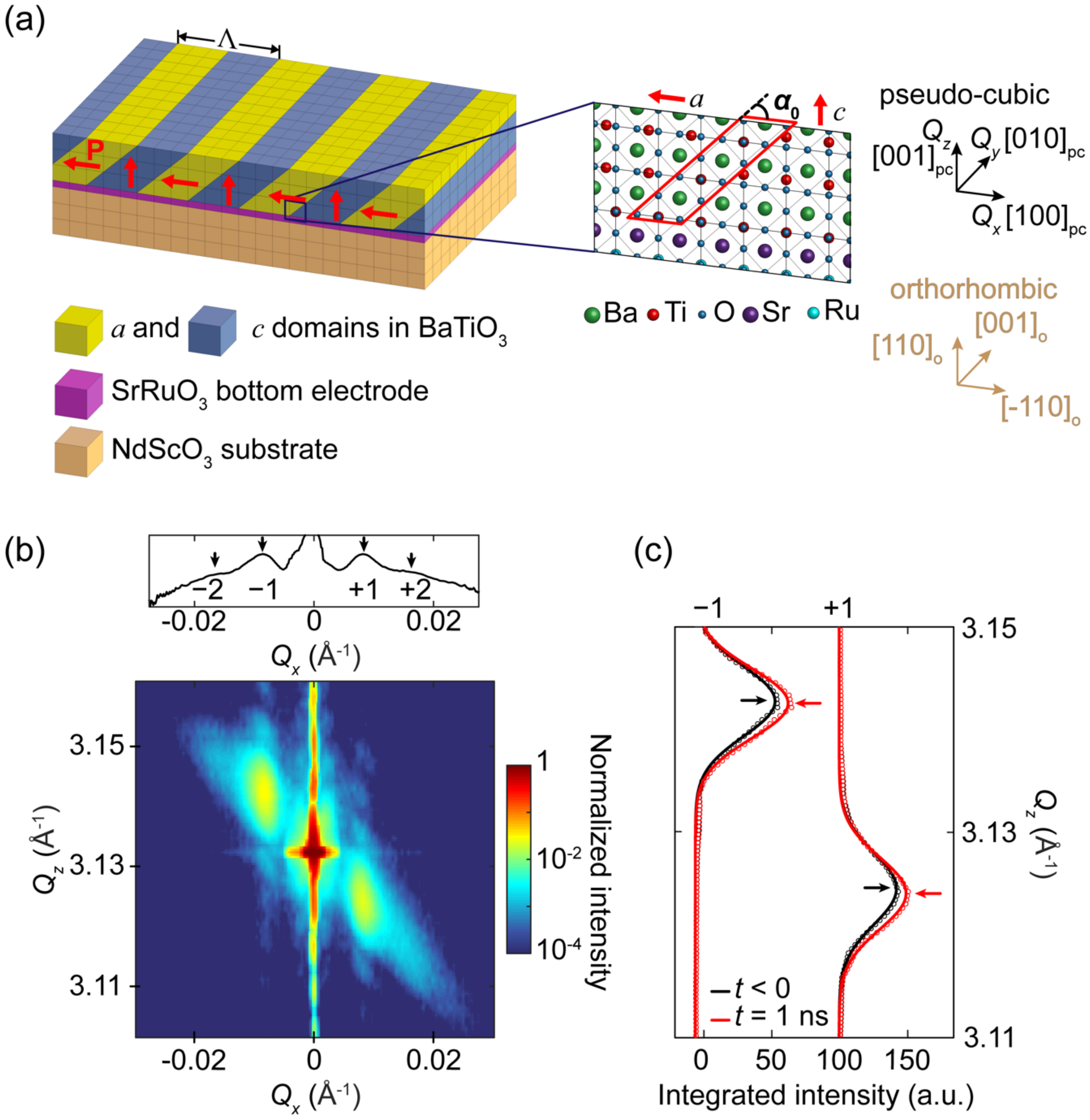

Ahn *et al*., Figure 2

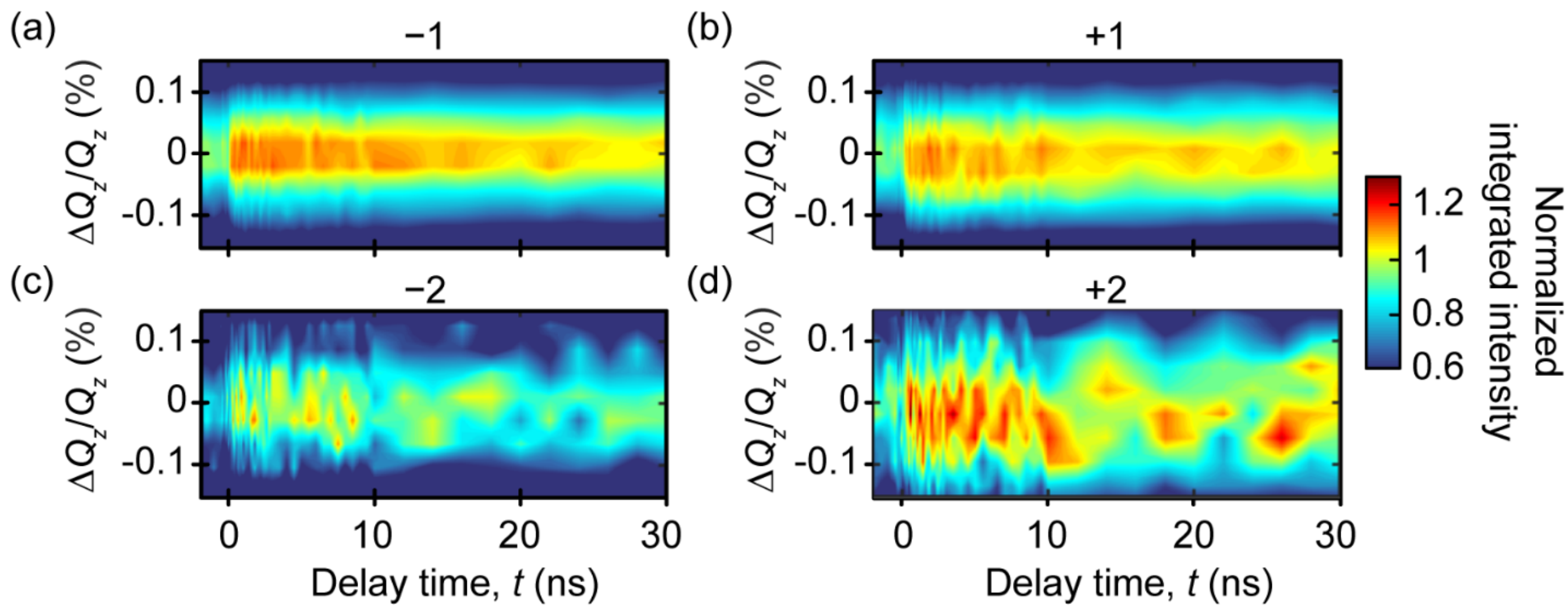

Ahn *et al*., Figure 3

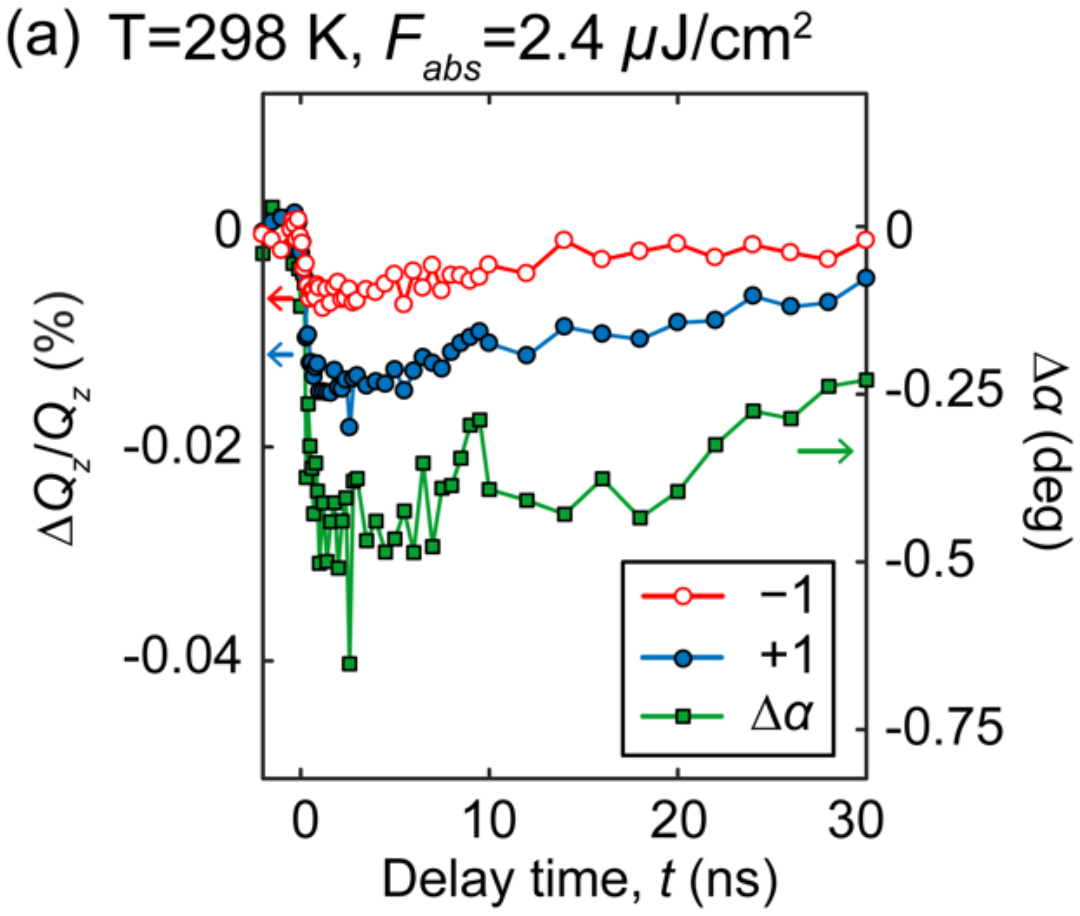

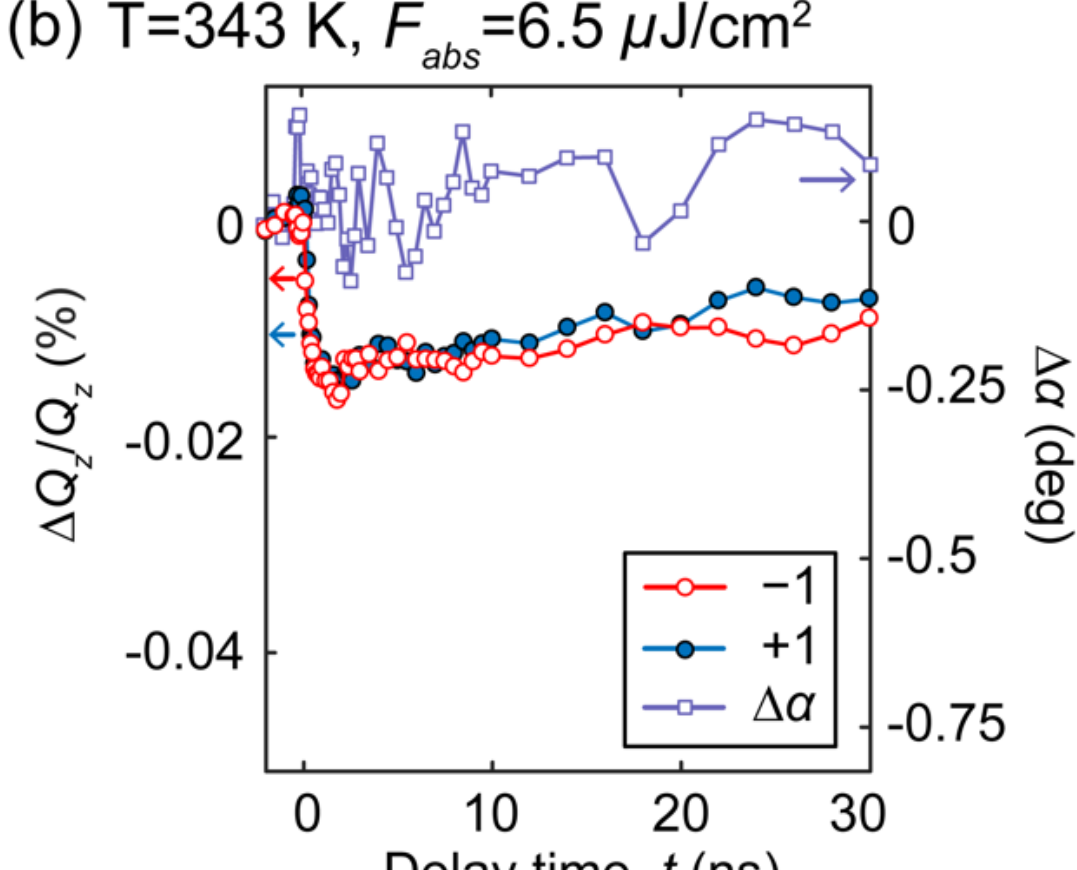

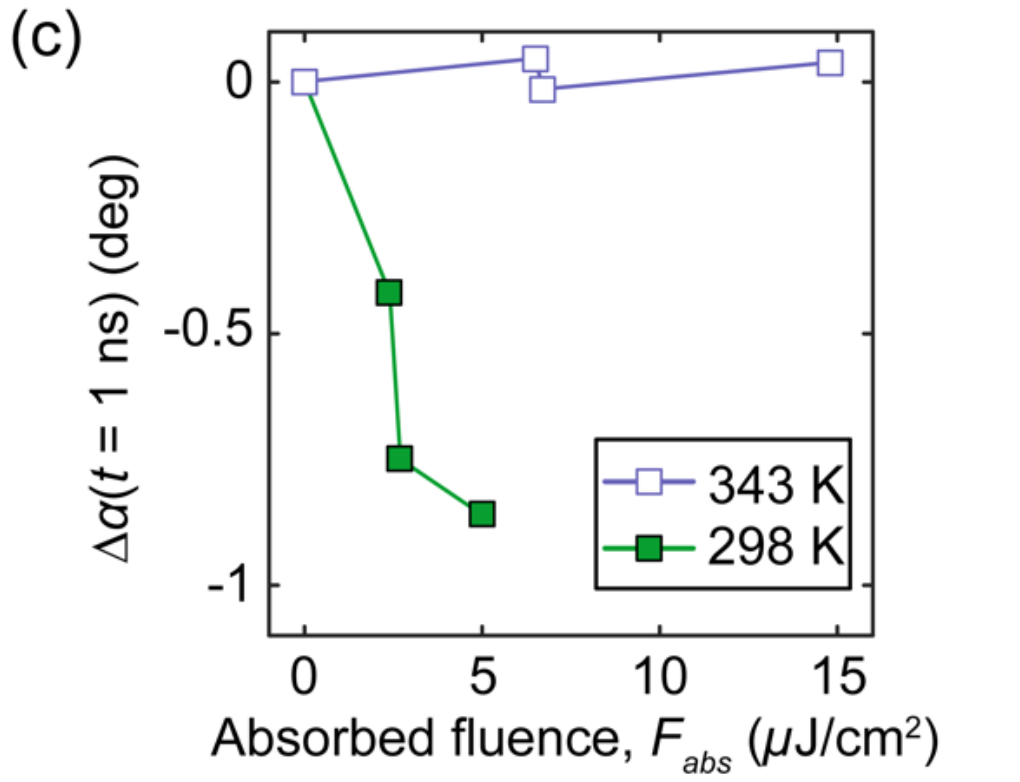

Ahn *et al*., Figure 4

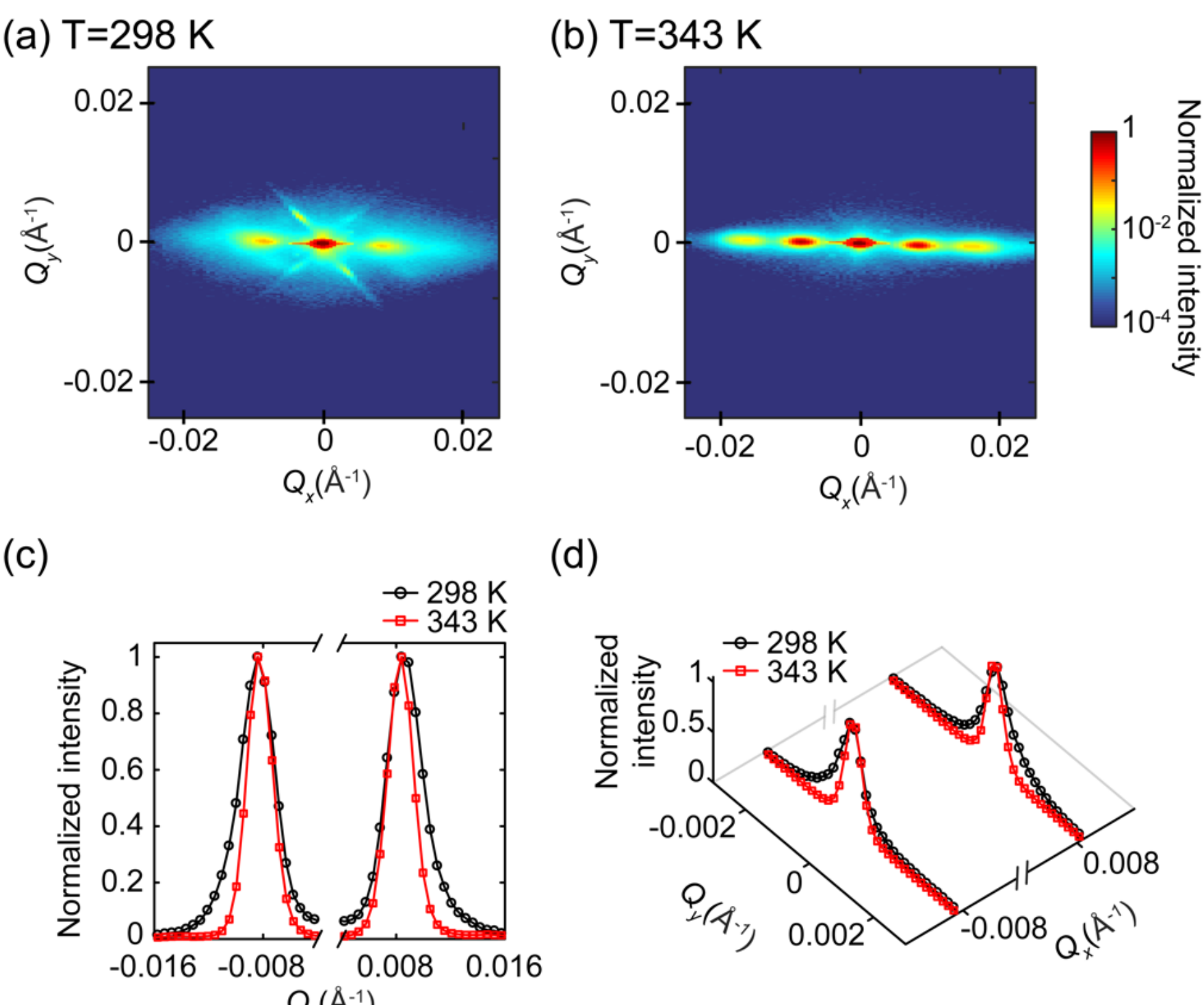

**Supplemental Material for "Dynamic Tilting of Ferroelectric Domain Walls Caused by Optically Induced Electronic Screening"**

Youngjun Ahn,[1] Hyeon Jun Lee,[1] Joonkyu Park,[1] Anastasios Pateras,[1,*] Arnoud S. Everhardt,[2] Silvia Damerio,[2] Tao Zhou,[3,†] Anthony D. DiChiara,[4] Haidan Wen,[4] Beatriz Noheda,[2,5] and Paul G. Evans[1,*]

[1] *Department of Materials Science and Engineering, University of Wisconsin-Madison, Madison, Wisconsin 53706, USA*

[2] *Zernike Institute for Advanced Materials, University of Groningen, 9747AG- Groningen, Netherlands*

[3] *ID01/ESRF, 71 Avenue des Martyrs, 38000 Grenoble Cedex, France*

[4] *Advanced Photon Source, Argonne National Laboratory, Argonne, Illinois 60439, USA*

[5] *CogniGron Center, University of Groningen, 9747AG- Groningen, Netherlands*

[*]pgevans@wisc.edu

## 1. Time-resolved x-ray diffraction

The time-resolved x-ray diffraction experiments were conducted at station 7-ID-C of the Advanced Photon Source. These measurements were conducted with x-ray incident angles near the BTO 002 Bragg condition, which is at $\theta_{In}$ = 20.06° at the 9 keV photon energy employed for these experiments. Diffracted x-ray photons were collected by a two-dimensional pixel array detector (Pilatus 100K, Dectris Ltd.) with a pixel size of 172 μm pixel size positioned at 1.53 m from the sample. The BTO film was excited by optical pulses with a central wavelength of 400 nm, 50 fs duration, and 1 kHz repetition rate provided by a Ti-Sapphire laser system. The optical incident angle was $\Delta\theta$ = 15° larger than the x-ray incident angle, as shown in Fig. S1. The optical pulses were s-polarized with respect to the plane of incidence.

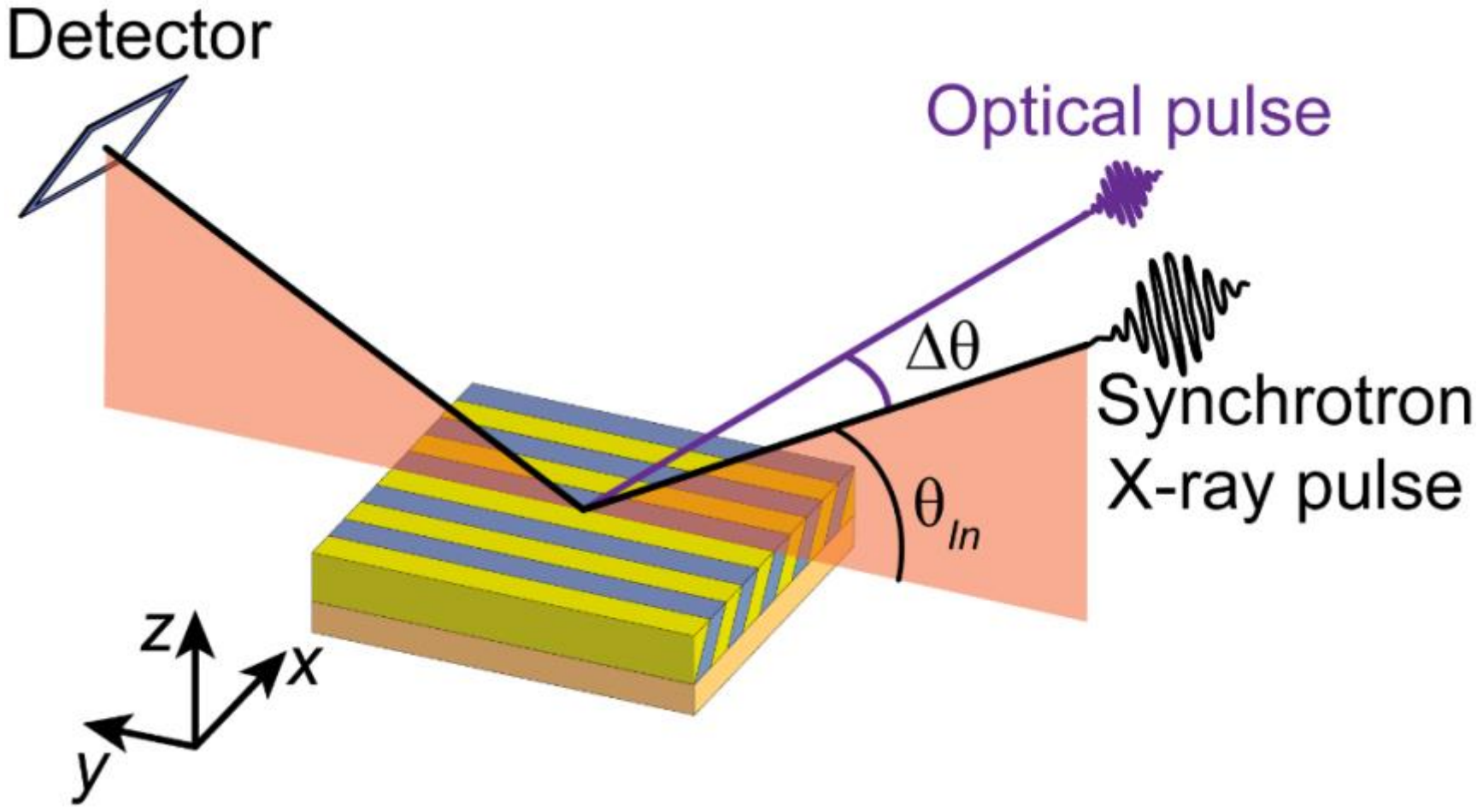

FIG. S1. Time-resolved synchrotron x-ray diffraction experiment.

## 2. Absorbed optical fluence and excitation density

The absorbed optical fluence $F_{abs}$ is:

$$F_{abs} = \frac{P_{in}}{Af}(1-R)\left[1-\exp\left(-\frac{(\epsilon+\epsilon_{NL})D}{\cos(90-(\theta_{In}+\Delta\theta))}\right)\right] \quad \text{(S1)}$$

Here $P_{in}$ is the measured optical power, $A$ is the full-width-at-half-maximum area of the laser spot on the sample surface, $f$ = 1 kHz is the optical repetition rate, $D$ = 78 nm is the BTO film thickness, $\theta_{In}+\Delta\theta$ is the incident angle of the laser with respect to the sample surface, and $\epsilon$ = 2.4×10$^3$ m$^{-1}$ is the linear optical absorption coefficient at a wavelength of 400 nm. The reflectance $R$=0.37 is calculated using the Fresnel equation.

The incident fluences given by $\frac{P_{in}}{Af}$ ranged from 4.3 mJ/cm$^2$ to 12.4 mJ/cm$^2$. Nonlinear absorption contributes significantly to the total optical absorption at these fluences and a wavelength of 400 nm [22]. We thus considered the nonlinear absorption by calculating $\epsilon_{NL}$ which is a function of the nonlinear absorption coefficient, optical pulse duration, and incident fluence [21]. The approach employed by Akamatsu *et al.* was used to calculate $\epsilon_{NL}$ [21]:

$$\epsilon_{NL} = \beta \frac{P_{in}}{Af} \frac{1}{\tau} (1 - R) \tag{S2}$$

Here, $\tau$ = 50 fs is the optical pulse duration and $\beta$ = 0.77 cm $GW^{-1}$ is the nonlinear absorption coefficient of BTO at 400 nm. For an incident fluence $\frac{P_{in}}{Af}$ = 4.3 mJ $cm^{-2}$, the value of $\epsilon_{NL}$ is $9.7 \times 10^3$ $m^{-1}$ and nonlinear absorption thus represents the largest contribution to the total absorption coefficient for the entire range of fluences reported here.

**3. Temperature dependence of the domain phase transformation and the domain wall tilt angle**

The temperature dependence of the domain phase transformation and the domain wall tilt angle $\alpha$ was studied using temperature-dependent x-ray diffraction at the European Synchrotron Radiation Facility. The x-ray photon energy was 8 keV. The diffracted x-ray intensity was measured using a pixel array detector positioned at 1.4 m from the sample.

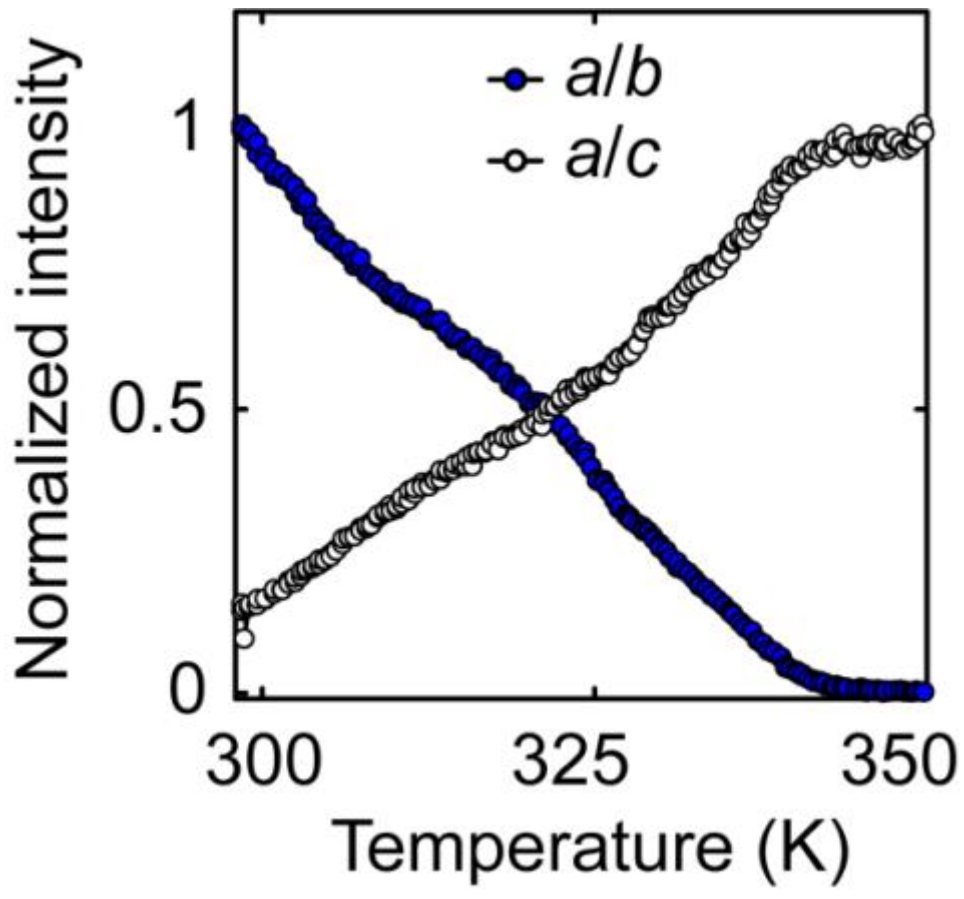

FIG. S2. Temperature dependence of the integrated intensity of *a*/*b* and *a*/*c* domain scattering.

The integrated intensities of *a*/*b* and *a*/*c* domain scattering were measured as a function of temperature from 298 K to 350 K as shown in Fig. S2. The intensity of *a*/*b* domain scattering decreased as the temperature increased, and disappeared above 340 K. The intensity of *a*/*c* domain scattering increased monotonically up to 340 K.

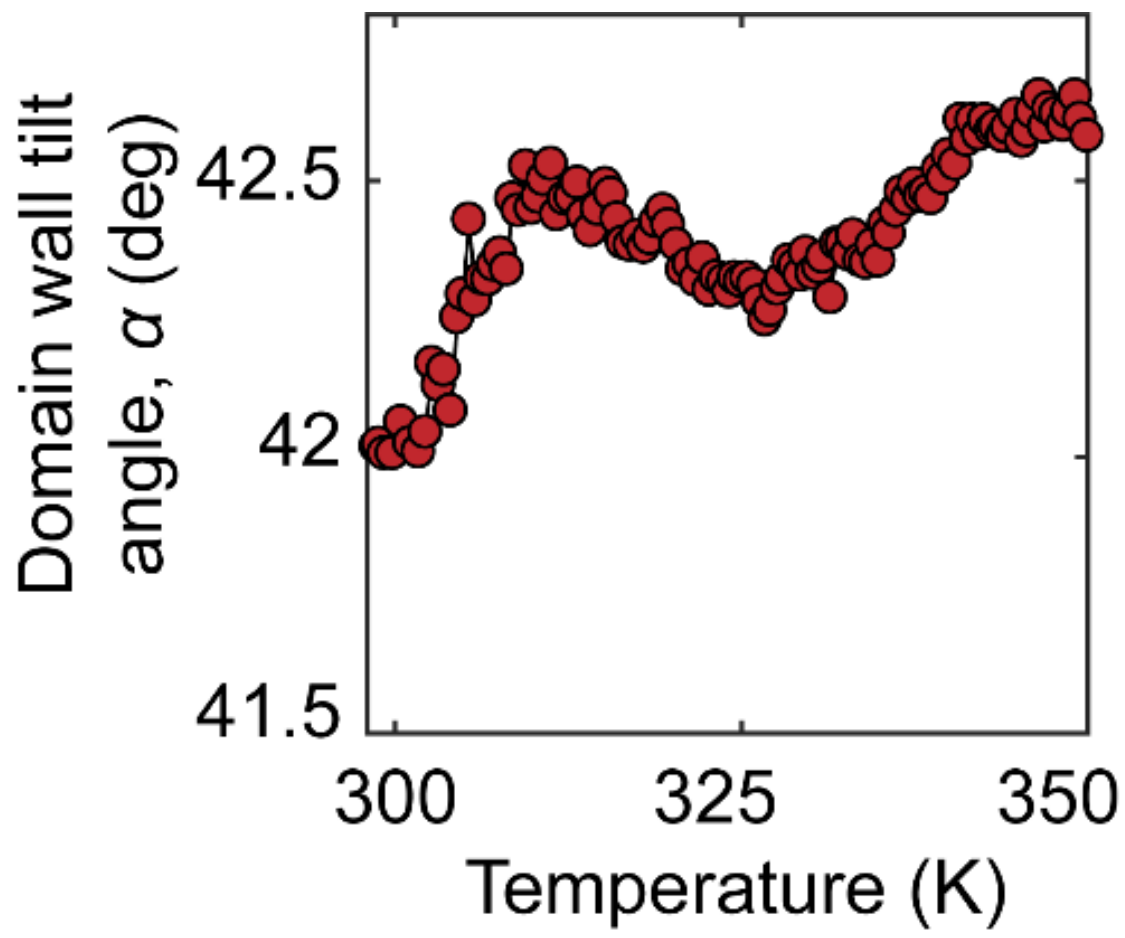

FIG. S3. Temperature dependence of the domain wall tilt angle, $\alpha$.

The temperature dependence of $\alpha$ from 298 K to 350 K is shown in Fig. S3. The domain wall tilt angle increases nonmonotonically. The nonmonotonic increasing trend of the domain wall tilt angle as a function of temperature is inconsistent with the optically induced change in the domain wall tilt angle, which exhibits a monotonically increasing trend. This result excludes laser-induced heating from the mechanism of the optically induced domain wall tilt.

**4. Thermodynamic description of the geometry and energetics of 90°-type ferroelastic domain walls**

Two alternative scenarios that can in principle lead to a change in the domain wall tilt angle: (i) a change in the lattice parameters along the out-of-plane direction, and (ii) a reduction of *a*-domain width. The sign and magnitude of the change in domain wall tilt angle due to these alternative mechanisms can be predicted using a free-energy model [23].

*4.a. Free-energy model*

The epitaxial strain and elastic compatibility at domain walls were described in the model by employing a continuous distribution of infinitesimal edge dislocations, in which the magnitude of elastic strain field is proportional to the dislocation density. The dislocation distributions at the film-substrate interfaces in the *a*-domain and *c*-domain are different because the strain due to the lattice misfit is different in the two domains. The elastic energy was computed using the resulting strain distribution and the elastic constants of BTO.

Three elastic energy terms associated with the domain wall tilt are taken into consideration [23,24]. The energy required to move the domain wall is equivalent to the energy required to create a screw dislocation along a pathway of the change in the domain-wall tilt in the fictitious dislocation method. The second term arises from the elastic interaction between the domain wall and the film-substrate interface in *c*-domain. The third contribution is the elastic interaction of the domain wall with the film-substrate interface in *a*-domain.

The elastic energy density $U_{elastic}$ due to the deviation of the equilibrium domain wall tilt angle $\alpha'$ from the ideal domain wall tilt angle $\alpha_{\text{ideal}} = 45°$ is [23]:

$$U_{elastic} = \frac{1}{2}k_1\varepsilon_T^2(\alpha' - \alpha_{\text{ideal}})^2 + k_2\varepsilon_c\varepsilon_T(\alpha' - \alpha_{\text{ideal}}) + k_2\varepsilon_a\varepsilon_T F\left(\frac{w}{D}\right)(\alpha' - \alpha_{\text{ideal}}) \quad \text{(S3)}$$

Here $\varepsilon_T$ is the tetragonality describing the elastic compatibility between *c*- and *a*-domains at the domain walls, $\varepsilon_c$ is the film-substrate misfit strain in the *c* domain, and $\varepsilon_a$ is the film-substrate misfit strain in the *a* domain, respectively. The constants $k_1$ and $k_2$ are given by $k_1 = (1+\ln 4)YD/(\pi\cdot(1-\nu^2))$ and $k_2 = YD/(1-\nu^2)$, where $Y$ and $\nu$ are the Young's modulus and Poisson ratio of BTO [25,26]. The function $F(w/D)$ describes interaction between the strain fields at the film-substrate interface in *a*-domain and at the domain wall. The value of $F(w/D)$ depends on the ratio of the $a$-domain width $w$ to the film thickness $D$ [23]. The value of $\alpha'$ calculated by minimizing $U_{elastic}$ with respect to $\alpha'$ is:

$$\alpha' = \frac{-k_2}{k_1 \varepsilon_T^2}\left(\varepsilon_c \varepsilon_T + \varepsilon_a \varepsilon_T F\left(\frac{w}{D}\right)\right) + \alpha_{\mathrm{ideal}} \tag{S4}$$

The change of the domain wall geometry is considered by investigating two reported optically induced responses: (i) an increase in the tetragonality $\varepsilon_T$ and (ii) a reduction of $a$-domain width $w$.

*4.b. Quantification of alternative domain-wall tilting models: elastic expansion and domain-width effects*

The change in the domain wall tilt angle $\Delta\alpha$ is obtained by $\Delta\alpha = \alpha'(\varepsilon_T, w) - \alpha_0$ where $\alpha'(\varepsilon_T, w)$ is the domain wall tilt angle obtained from Eq. (S4) when $\varepsilon_T$ and $w$ are varied. $\alpha_0$ is the equilibrium domain wall tilt angle of the epitaxially strain BTO film on the NSO substrate. The lattice parameters of bulk BTO and pseudo-cubic NSO are used to calculate $\Delta\alpha$ in Eq. (S4) [27,28].

The optically induced expansion causes a change in tetragonality by 0.01% for $F_{abs}$ = 2.4 $\mu$J/cm$^2$. The calculated value of the $\Delta\alpha$ due to the change in tetragonality by 0.01% is 0.001° toward the substrate normal. The observed tilt under these conditions is −0.5°. The tilt predicted by the change in tetragonality is thus two orders of magnitude smaller than experimentally observed value and has the opposite direction of the measured tilt.

The effect of the domain width on the tilt angle was investigated by computing the value of $\Delta\alpha$ as a function of $w$ from 0 to 78 nm. The reduction of the $a$-domain width from its equilibrium width of 39 nm leads to a decrease of the value of $F(w/D)$ [23]. The calculated value of $\Delta\alpha$ as a function of $w$ is shown in Fig. S4. A produces a smaller change by an order of magnitude than the observed maximum $\Delta\alpha$, even when the *a* domain completely disappears. Further, the reduction of *a*-domain width following optical excitation was observed in the single-phase *a*/*c* domain pattern [21], and thus the domain-wall tilting due to the decrease in *a*-domain width is expected to occur at both 298 K and 343 K domain phases. These results suggest that the optically induced structural changes are not responsible for the observed change in the domain wall tilt angle.

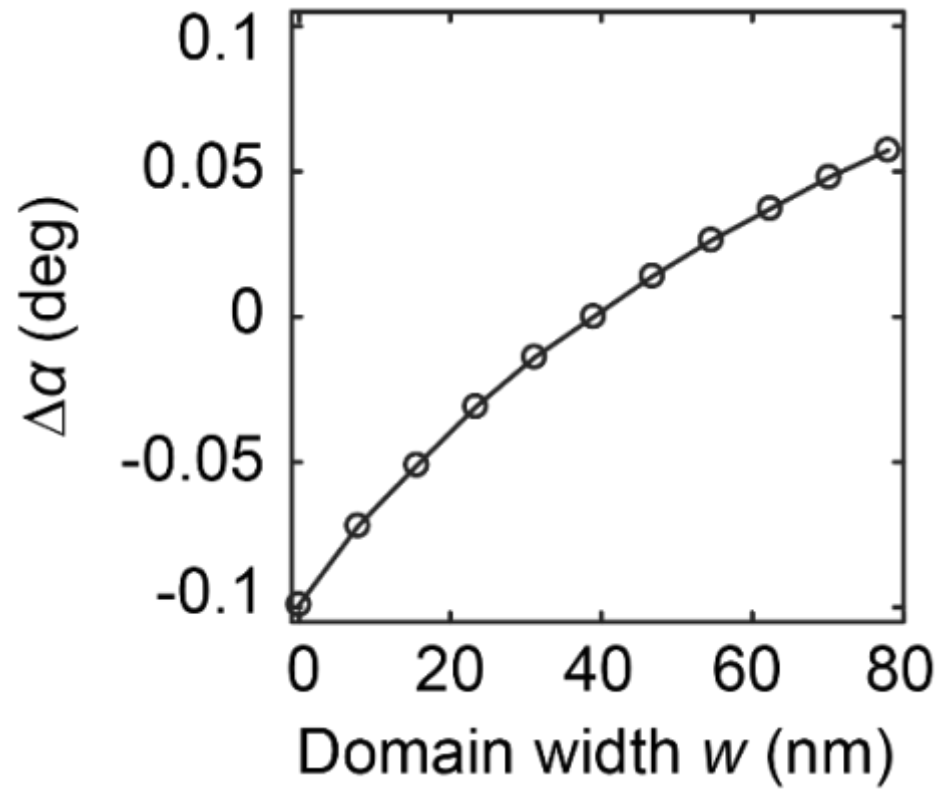

FIG. S4. Change in domain-wall tilt angle $\Delta\alpha$ as a function of *a*-domain width *w*.

**5. Temperature dependence of intensity distribution in $Q_x$-$Q_z$ section**

The better order in the high-temperature phase is apparent in the distribution of scattered x-ray intensity in the $Q_x$-$Q_z$ section of reciprocal space near the 002 BTO reflection. The location of this section is illustrated in Fig. S5(a). The diffraction pattern at 298 K is shown in Fig. S5(b), repeated from Fig. 1(b). At 343 K as in Fig. S5(c), the higher-order domain scattering intensity maxima are sharper and exhibit higher intensity. In addition, the BTO layer thickness fringes of the ±1 and ±2 orders exhibit higher intensity at 343 K than at 298 K.

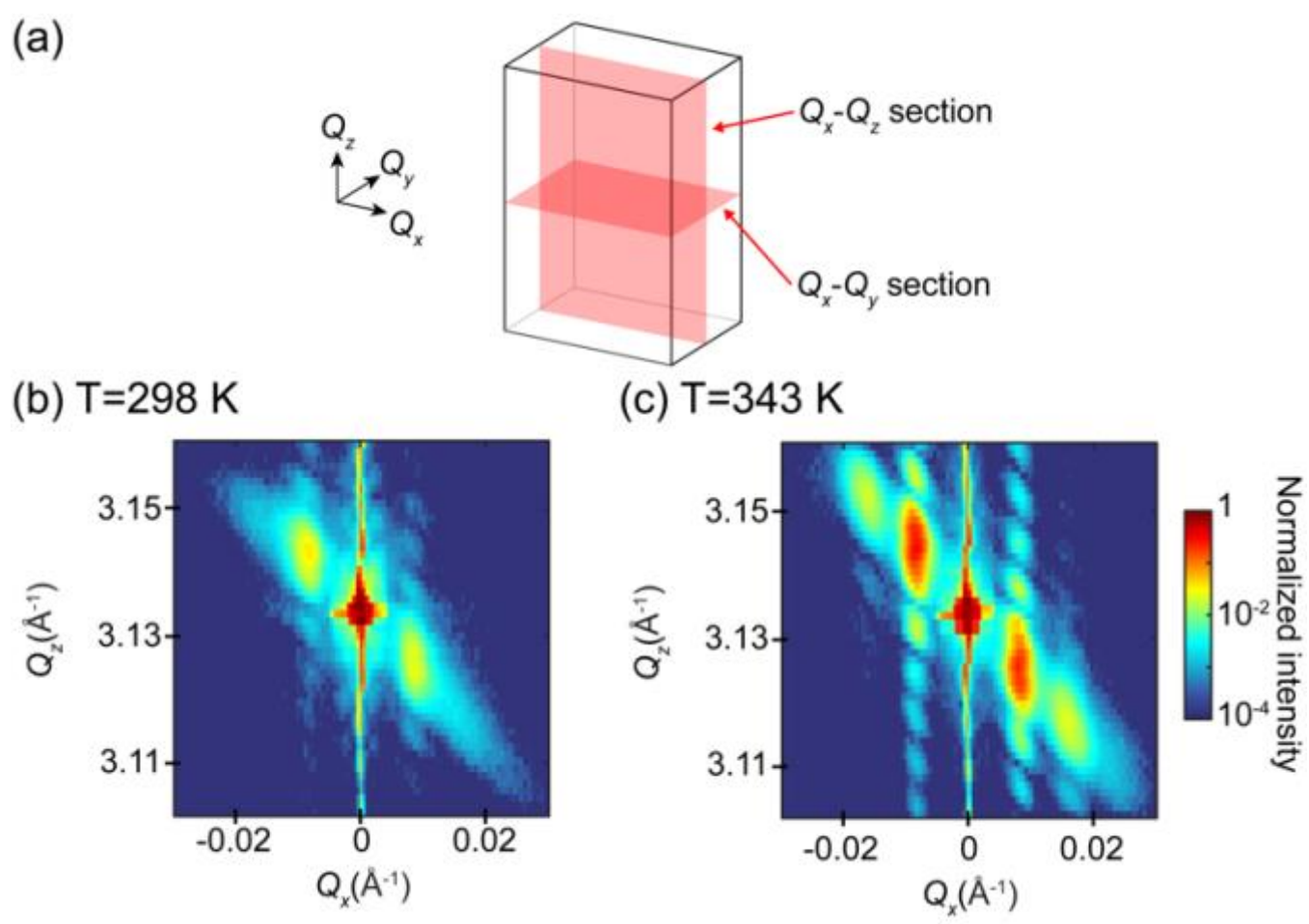

FIG. S5. (a) Schematic of $Q_x$-$Q_z$ and $Q_x$-$Q_y$ sections of reciprocal space. Diffracted x-ray intensity distributions in the $Q_x$-$Q_z$ cross sections at (b) 298 K and (c) 343 K.

**6. Fluence dependence of tilt dynamics**

The dynamics of the change in the domain wall tilt angle and recovery are shown in Fig. S6 for several values of $F_{abs}$ at 298 K.

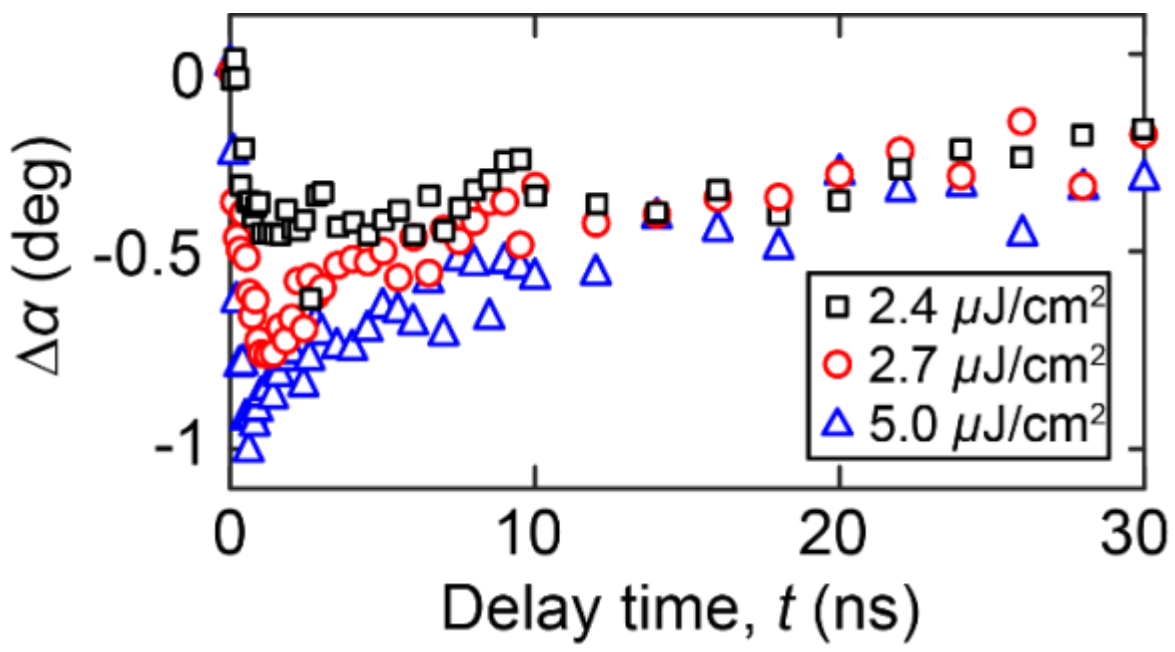

FIG. S6. Time dependence of the change in domain wall tilt angle $\Delta\alpha$ for $F_{abs}$ = 2.4 $\mu$J/cm$^2$, 2.7 $\mu$J/cm$^2$, and 5.0 $\mu$J/cm$^2$.